\documentclass[10pt,twocolumn,aps,prl,longbibliography,nofootinbib,floatfix]{revtex4-2}
\usepackage{graphicx,bm,xcolor, bbold}
\usepackage[normalem]{ulem}
\usepackage{hyperref}
\hypersetup{colorlinks, linkcolor={blue},citecolor={blue},urlcolor={blue}}
\usepackage{graphicx}
\usepackage{amsmath,amssymb,amsfonts}
\usepackage{bm}
\usepackage{color}
\graphicspath{fig}
\usepackage{epstopdf}
\usepackage{multirow}
\usepackage{scalerel}

\newcommand{\BZ}{\text{BZ}}

\newcommand{\vect}[1]{\bm{\mathrm{#1}}}

\newcommand{\bv}{\vect{b}}
\newcommand{\jv}{\vect{j}}
\newcommand{\kv}{\vect{k}}
\newcommand{\nablav}{\vect{\nabla}}
\newcommand{\pv}{\vect{p}}

\newcommand{\rv}{\vect{r}}

\newcommand{\vv}{\vect{v}}

\newcommand{\calH}{\mathcal{H}}

\newcommand{\nv}{\vect{n}}

\begin{document}
\title{Superfluid density in linear response theory:\\pulsar glitches from the inner crust of neutron stars}
\author{Giorgio Almirante}
\email{giorgio.almirante@ijclab.in2p3.fr}
\affiliation{Universit\'e Paris-Saclay, CNRS/IN2P3, IJCLab, 91405 Orsay, France}
\author{Michael Urban}
\email{michael.urban@ijclab.in2p3.fr}
\affiliation{Universit\'e Paris-Saclay, CNRS/IN2P3, IJCLab, 91405 Orsay, France}
\begin{abstract}
The question of whether there are enough superfluid neutrons in the inner crust of neutron stars to explain pulsar glitches remains a topic of debate. 
Previous band structure calculations suggest that the entrainment effect significantly reduces the superfluid density.
In this letter, a new derivation of the BCS expression for the superfluid density is given. 
We compute it in the superfluid band theory framework through linear response theory, for a small relative velocity between superfluid and normal components, under the assumption that the pairing gap in the rest frame of the superfluid is constant and not affected by the perturbation.
Our result suggests that a formula extensively used in neutron star physics is incomplete. Numerical evaluations for two realistic configurations reveal that the previously neglected contribution drastically alters the picture of the superfluid reservoir in the inner crust of neutron stars, suggesting that about $90\%$ of the neutrons are effectively superfluid.
\end{abstract}
\maketitle
\textit{Introduction}---It is generally believed that the inner crust of neutron stars contains a crystal lattice of nuclei (clusters) and a superfluid gas of unbound (dripped) neutrons \cite{Chamel08}. 
The superfluid neutrons play a crucial role, e.g., in the understanding of pulsar glitches \cite{Anderson75}. 
Under the assumption that all unbound neutrons in the inner crust are superfluid, their moment of inertia is sufficient to explain the glitches as a sudden transfer of angular momentum from the superfluid neutrons to the lattice. 
However, as pointed out in \cite{Prix02}, there can be a so-called ``entrainment'' effect, which causes a part of the unbound neutrons to flow with the nuclei, effectively reducing the number of superfluid neutrons. 
A reduction of the superfluid density is actually not limited to the inner crust of neutron stars, it happens whenever a superfluid system is non-uniform \cite{Leggett98}.

Quantitative calculations of the entrainment effect started with the pioneering work by Carter, Chamel, and Haensel \cite{Carter05,Carter05A,Chamel06} in the framework of band theory for neutrons \cite{Chamel05}, analogous to the band theory for electrons in solids \cite{Ashcroft}.
These calculations were subsequently refined and carried out for the entire inner crust \cite{Chamel12A}.
The results of these calculations indicate that the entrainment can be very strong, reducing drastically the superfluid density in the inner crust. 
This is in contradiction with the observed glitches of certain pulsars \cite{Chamel13}, unless one gives up the common belief that only the crust is responsible for the glitches \cite{Andersson12}.
Furthermore, the entrainment affects also strongly the speed of the transverse lattice phonons and hence the heat capacity and heat conductivity of the inner crust \cite{Chamel13B}.

However, what was actually computed in ``normal'' band theory \cite{Chamel05,Chamel06,Chamel12A,Kashiwaba19,Sekizawa22} is the conduction density in the normal phase, which coincides with the superfluid density in the limit of zero pairing gap ($\Delta\to 0$).
Inclusion of pairing was addressed in \cite{Carter05A,Chamel25} within the BCS approximation.
There the authors conclude that pairing should not have any significant effect on the superfluid density.
But this is in contradiction with \cite{Watanabe17}, where Hartree-Fock-Bogoliubov (HFB) calculations\footnote{The HFB equations are also called Bogoliubov-de-Gennes equations in condensed matter physics.} were performed in a one-dimensional sinusoidal lattice potential, and it was found that the effects of band structure are suppressed when the pairing gap $\Delta$ is of order or greater than
the strength of the lattice potential.
In \cite{Martin16}, the superfluid density was computed in the framework of superfluid hydrodynamics, which is valid in the limit of strong pairing (i.e., if the coherence length of the Cooper pairs is small compared to the size of the inhomogeneities, which is probably not the case in the inner crust), and the resulting superfluid density was found to be larger than 90\% of the total neutron density in almost the entire inner crust except the outermost region.
In \cite{Minami22}, it was shown that the BCS approximation can lead to an underestimation of the superfluid density.
To go beyond the BCS approximation, the first fully self-consistent HFB calculations were performed for the slab (``lasagna'') phase \cite{Almirante24,Yoshimura24}. 
In the case of the rod (``spaghetti'') phase \cite{Almirante24A}, we found that the HFB calculation reproduces the result of normal band theory \cite{Carter05} in the limit $\Delta\to0$, but then the superfluid density increases rapidly with $\Delta$ and reaches values similar to those obtained in superfluid hydrodynamics when the pairing gap takes realistic values.

In this letter, we solve this issue by deriving the expression for the superfluid density in linear response theory. 
We find an additional contribution that was neglected in \cite{Carter05A,Chamel25}. Evaluating numerically this contribution for the crystal phase of the inner crust of neutron stars, we find a strong dependence on the value of the pairing gap, resulting in much larger superfluid densities at realistic values of the gap. 

\textit{Superfluid density}---In homogeneous one-component systems at zero temperature, the superfluid density\footnote{As common in the nuclear physics literature, we use the term ``density'' and the symbol $\rho$ for the number density and not for the mass density.} $\rho_S$ is always equal to the total density $\rho$ \cite{Leggett98} (under the condition that the flow has velocity smaller than the Landau velocity, where superfluidity starts to be destroyed).
However, if the system is not homogeneous or at finite temperature, the superfluid density $\rho_S$ is smaller than the average density $\langle\rho\rangle$, and the system behaves as if a superfluid component coexisted with a normal fluid.
To avoid unnecessary complications of the discussion, we assume that the interactions within the system have no momentum dependence, so that the microscopic effective mass $m^*$ is equal to the bare mass $m$ and the momentum density $\jv$ is equal to the mass current.
Then the superfluid density $\rho_S$ can be defined via the coarse-grained current
\begin{equation} \label{eq:andreevbashkin}
  \langle\jv\rangle =(\langle\rho\rangle -\rho_S) m\vv +  \rho_S m\vv_S \,,
\end{equation}
where $\vv$ is the velocity of the normal fluid, and $\vv_S$ is the so-called superfluid velocity $\vv_S = \nablav\phi/(2m)$, where $\phi$ is the coarse-grained phase of the pairing field $\Delta$ \cite{Pethick10}.

In the reference frame in which the superfluid component is at rest, i.e. $\vv_S=0$, Eq.~(\ref{eq:andreevbashkin}) takes the form
\begin{equation} \label{eq:formalcurrent}
    \langle\jv\rangle = (\langle\rho\rangle -\rho_S) m\vv \,.
\end{equation}
As explained in \cite{Almirante24,Almirante24A}, this physical situation can be described within the time independent Hartree-Fock-Bogoliubov (HFB) framework.
If the mean-field Hamiltonian in the rest frame of the normal component reads $h^{(0)}=\pv^2/(2m)+U(\rv)-\mu$, where $U$ is the periodic mean field and $\mu$ the chemical potential, the system in which the normal component moves with velocity $\vv$ is described by $h^{(\vv)}=h^{(0)}-\vv\cdot\pv$ \cite{LandauLifshitz3}. 
Hence, the term $-\vv\cdot\pv$ will appear as a perturbation in the HFB matrix $\calH^{(\vv)}$ (see below).
Notice that, in order to have $\vv_S=0$, one has to require that the phase of the pairing gap $\Delta$ is constant in average.
This is of course fulfilled if not only $U$ but also $\Delta$ is periodic \cite{Almirante24,Almirante24A}.

\textit{Band theory}---In periodic systems, the momentum space labels can be decomposed into integer multiples of the primitive vectors of the reciprocal lattice (i.e., of $2\pi/L$ in the case of a simple cubic lattice with lattice spacing $L$), $n_1\bv_1+n_2\bv_2+n_3\bv_3\equiv\nv$, and a continuous momentum $\kv$ known as Bloch momentum, which is limited to the first Brillouin zone (BZ). We denote the corresponding momentum eigenstates as $|\nv\kv\rangle$ (one can think of them in coordinate space as $\langle\rv|\nv\kv\rangle\propto e^{i(\nv+\kv)\cdot\rv}$). Then the Bloch theorem ensures that the Hamiltonian $h^{(\vv)}$ is diagonal in $\kv$ \cite{Ashcroft}.

In the Hartree-Fock (HF) theory one can compute the band structure, solving for each $\kv\in\text{BZ}$ the eigenvalue problem for the mean-field Hamiltonian $h^{(0)}$, namely
\begin{equation} \label{eq:HFH}
    h^{(0)}|\alpha\kv\rangle=\xi_{\alpha\kv}|\alpha\kv\rangle \,,
\end{equation}
where $|\alpha\kv\rangle$ is the eigenstate with single-particle energy $\xi_{\alpha\kv}$. For better readability, we will from now drop the label $\kv$ and simply write $|\nv\rangle$, $|\alpha\rangle$, $\xi_{\alpha}$, etc., but one has to keep in mind that every quantity depends on the same $\kv$.

\textit{BCS approximation}---What in nuclear structure theory is called the BCS approximation is the assumption that the pairing gap $\Delta$ is diagonal in the HF basis.
In fact, we simplify the problem further by taking the pairing gap constant, as it was also done, e.g., in \cite{Chamel25}. Then, the eigenvalue problem for the HFB matrix $\calH_{\alpha\beta}^{(\vv)}$ written in the HF basis becomes 
\begin{equation} \label{eq:BCSH}
\resizebox{\columnwidth}{\height}{
 $\displaystyle
 \sum_\beta
 \begin{pmatrix}
     \xi_\alpha\delta_{\alpha\beta}-\vv\cdot\pv_{\alpha\beta} & \Delta\delta_{\alpha\beta} \\
     \Delta\delta_{\alpha\beta} & -\xi_\alpha\delta_{\alpha\beta}-\vv\cdot\pv_{\alpha\beta}
 \end{pmatrix}\! 
 \begin{pmatrix}u_\beta\\v_\beta\end{pmatrix}
 = E_\alpha\! \begin{pmatrix}u_\alpha\\v_\alpha\end{pmatrix}\!,
 $}
\end{equation}
where $\pv_{\alpha\beta}=\langle\alpha|\pv|\beta\rangle$. In the absence of the term $-\vv\cdot\pv$, i.e. taking $\calH_{\alpha\beta}^{(0)}$,
one gets the usual BCS expressions for quasi-particle energies $E_\alpha$ and eigenvector components
\begin{equation} \label{eq:BCSEv}
    E_\alpha=\sqrt{\xi^2_\alpha+\Delta^2}\,,\quad v^2_\alpha=\frac{1}{2}-\frac{\xi_\alpha}{2E_\alpha}\,, \quad u^2_\alpha=\frac{1}{2}+\frac{\xi_\alpha}{2E_\alpha} \,.
\end{equation}
Notice that the occupation numbers $v^2_\alpha$ are just the elements of the density matrix $\rho_{\alpha\beta}=v^2_\alpha\delta_{\alpha\beta}$ in the HF basis.

\textit{Linear response}---Since we want to treat the $-\vv\cdot\pv$ term as a perturbation, we will refer to the quantitities given in Eq.~(\ref{eq:BCSEv}) as the unperturbed ones. In the unperturbed system, there is no current, thus the current of the perturbed system comes just from the perturbative correction $\rho'_{\alpha\beta}$ to the density matrix, i.e.,
\begin{equation} \label{eq:current}
     \langle\jv\rangle = 2 \int_{\BZ} \frac{d^3k}{(2\pi)^3} \sum_{\alpha\beta} \rho'_{\alpha\beta}\pv_{\beta\alpha} \,.
\end{equation}
(The factor $2$ accounts for the spin degeneracy).

The next task is to compute the correction $\rho'_{\alpha\beta}$. Migdal \cite{Migdal59} encountered a very similar problem when he studied the moment of inertia of a superfluid nucleus. Namely, he considered the linear response of a deformed nucleus to the perturbation $V=\Omega L_z$, $\Omega$ being the rotation frequency and $L_z$ the $z$ component of the angular momentum $\vect{r}\times\pv$. In order to compute the perturbative correction $\rho'_{\alpha\beta}$, one can perform perturbation theory on the HFB equation \eqref{eq:BCSH} (see appendix). Taking the linear order in a perturbation $V$ that is odd in $\pv$, such that it appears with the same sign in the upper left and the lower right elements of $\calH_{\alpha\beta}$ (such as $\Omega L_z$ or $\vv\cdot\pv$), one gets Migdal's result \cite{Migdal59}
\begin{equation} \label{eq:migdal}
    \rho'_{\alpha\beta} = \frac{\xi_\alpha \xi_\beta - E_\alpha E_\beta + \Delta^2}{2 E_\alpha E_\beta (E_\alpha + E_\beta)} V_{\alpha\beta} \,,    
\end{equation}
where we are neglecting the effect of the perturbation on the pairing gap.

Now we can rewrite explicitly the current in Eq.~(\ref{eq:current}) for the perturbation $V=-\vv\cdot\pv$.
Plugging Eq.~(\ref{eq:migdal}) into Eq.~(\ref{eq:current}), one gets ($i,j=x,y,z$) 
\begin{equation}\label{eq:responseRS}
    \langle j^i\rangle = \sum_{ij} \big(R^{ij} - S^{ij}\big) v^j\,,
\end{equation}
where $R$ and $S$ are tensors given by
\begin{align}\label{eq:responseR}
    R^{ij} =& \int_{\BZ}\frac{d^3k}{(2\pi)^3} \sum_{\alpha\beta}
    \frac{E_\alpha E_\beta - \xi_\alpha \xi_\beta + \Delta^2}{E_\alpha E_\beta (E_\alpha + E_\beta)}
    p_{\alpha\beta}^i p_{\beta\alpha}^j\,,\\
    \label{eq:responseS}
    S^{ij} =& \int_{\BZ}\frac{d^3k}{(2\pi)^3}\sum_{\alpha\beta} \frac{2\Delta^2}{E_\alpha E_\beta (E_\alpha + E_\beta)}
    p_{\alpha\beta}^i p_{\beta\alpha}^j\,.
\end{align}
Using that the pairing gap is constant, i.e. $\Delta^2=E^2_\alpha-\xi^2_\alpha=E^2_\beta-\xi^2_\beta$, Eq.~\eqref{eq:responseR} can be rewritten as
\begin{equation}\label{eq:responseR-rewritten}
   \resizebox{\columnwidth}{\height}{$\displaystyle
    R^{ij} =-2\!\int_{\BZ}\frac{d^3k}{(2\pi)^3}
    \Bigg[\sum_{\alpha} 
    \frac{\partial v^2_\alpha}{\partial \xi_\alpha}
    p^i_{\alpha\alpha} p^j_{\alpha\alpha}
    +
    \sum_{\alpha\neq\beta} 
    \frac{v^2_\alpha-v^2_\beta}{\xi_\alpha-\xi_\beta}
    p^i_{\alpha\beta} p^j_{\beta\alpha}\Bigg].$}
\end{equation}
In order to proceed further, one needs to express the momentum operator $\pv$ in the HF basis.
Working in momentum space, we can do this by inserting a complete set of momentum eigenstates, which gives $\pv_{\alpha\beta} = \sum_{\nv}(\nv+\kv)\langle\alpha|\nv\rangle\langle\nv|\beta\rangle$. However, in order to simplify Eq.~\eqref{eq:responseR-rewritten} analytically, it is better to use the following trick.
Computing the derivative of $\langle\alpha|h^{(0)}|\beta\rangle=\xi_\alpha\delta_{\alpha\beta}$ with respect to the Bloch momentum $\kv$, and using the fact that $\langle\alpha|\beta\rangle=\delta_{\alpha\beta}$ is independent of $\kv$,
 one gets
\begin{equation} \label{eq:momentum}
    \frac{\pv_{\alpha\beta}}{m}=
    \frac{\partial\xi_\alpha}{\partial \kv} \delta_{\alpha\beta} +
    (\xi_\alpha-\xi_\beta)\sum_{\nv}\frac{\partial\langle\alpha|\nv\rangle}{\partial \kv}\langle\nv|\beta\rangle.
\end{equation}
Then inserting Eq.~(\ref{eq:momentum}) into Eq.~\eqref{eq:responseR-rewritten}, and performing some algebraic calculations (see appendix), one can show that
\begin{equation}\label{eq:responseR-final}
  R^{ij} = 2 m\int_{\BZ}\frac{d^3k}{(2\pi)^3}\sum_{\alpha} v_\alpha^2\, \delta^{ij} = m\langle\rho\rangle\delta^{ij} \,.
\end{equation}
Comparing Eq.~\eqref{eq:responseRS} with the expression for the momentum density in Eq.~(\ref{eq:formalcurrent}), this implies that the superfluid density $\rho_S$ can be written in tensor form as
\begin{equation}
    \rho_S^{ij} = \frac{S^{ij}}{m} \,.
\end{equation}
Taking separately the diagonal ($\alpha=\beta$) and off-diagonal ($\alpha\ne\beta$) contributions to the sum in Eq.~\eqref{eq:responseS}, one can rewrite the above expression as
\begin{align} \label{eq:supdens}
    \rho_S^{ij} = & \int_{\BZ}\frac{d^3k}{(2\pi)^3} \Bigg[m \sum_{\alpha} \frac{\Delta^2}{E_\alpha^3} \frac{\partial\xi_{\alpha}}{\partial k^i}\frac{\partial\xi_{\alpha}}{\partial k^j} \nonumber \\
    & +\frac{2}{m}\sum_{\alpha\neq\beta} \frac{\Delta^2}{E_\alpha E_\beta (E_\alpha + E_\beta)} 
    p^i_{\alpha\beta} p^j_{\beta\alpha}
    \Bigg] \,.
\end{align}
Notice that, if the system has cubic symmetry, one has $\rho_S^{ij}=\rho_S \delta^{ij}$ \cite{Carter06} and the scalar superfluid density is $\rho_S=\sum_i\rho_S^{ii}/3$.

In the diagonal ($\alpha=\beta$) contribution (first term), one can recognize the formula for the superfluid density as derived in \cite{Carter05A,Chamel25}.
In the limit $\Delta\rightarrow 0$, the factor $\Delta^2/E_\alpha^3$ tends to $2\delta(\xi_\alpha)$, so that this contribution tends to the conduction density in the normal phase, $\rho_c$ \cite{Chamel12A}, while the contribution of the sum over $\alpha\ne\beta$  (second term) tends to zero.
But as soon as $\Delta$ is comparable with the spacing between bands, the second term can no longer be neglected.
This makes many of the results in the existing literature \cite{Carter05,Chamel05,Chamel12A} correct only in the limit $\Delta\to 0$, but not for realistic values of $\Delta\neq 0$ (as shown also in \cite{Almirante24A}).

\textit{Neutron star crust}---Now we evaluate numerically Eq.~(\ref{eq:supdens}) to compute the superfluid density $\rho_S$ in the crystal phase of the inner crust of neutron stars. 
We take the mean-field potentials from Oyamatsu \cite{Oyamatsu93,Oyamatsu94}, at baryon densities $\rho_b=0.030$~fm$^{-3}$ and $\rho_b=0.055$~fm$^{-3}$ (what we call $\rho$ in this letter refers only to the neutron density), in the simple cubic lattice with spacing $L=34.75$~fm and $L=27.99$~fm, respectively. 
Then we solve the HF equations \eqref{eq:HFH} with Bloch boundary conditions, using the discrete Fourier transform to the momentum basis with $22^3$ points in the cubic cell, and $24^3$ points in the first Brillouin zone. 
We can thus compute the corresponding BCS expressions in Eq.~(\ref{eq:BCSEv}) and also the matrix elements of the momentum operator $\pv_{\alpha\beta}$ as said above Eq.~\eqref{eq:momentum}.
With these ingredients, one can use Eq.~(\ref{eq:supdens}) to compute the superfluid density $\rho_S$, at different values of the pairing gap $\Delta$. 

\begin{figure}
    \centering
    \includegraphics{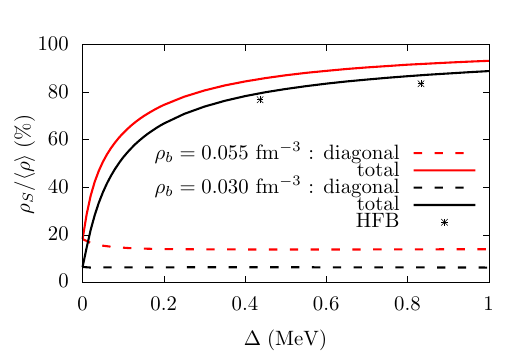}
    \caption{Superfluid fraction $\rho_S/\langle\rho\rangle$ for two different baryon densities $\rho_b$ (using the mean field potentials of \cite{Oyamatsu93,Oyamatsu94}) computed for different values of the pairing gap $\Delta$.
    Solid lines are the results from Eq.~(\ref{eq:supdens}).
    Dashed lines are instead just the contribution of the diagonal elements ($\alpha=\beta$) to the sum in Eq.~(\ref{eq:supdens}).
    Black stars are HFB results obtained performing the self-consistency only in the pairing channel with reduced coupling constant (see text). Data files available at \cite{Almirante2025data}.}
    \label{fig:supdens}
\end{figure}

Our results are collected in Fig.~\ref{fig:supdens}.
We show both the total superfluid density $\rho_S$ (solid lines) and the contribution resulting just from the sum over diagonal elements ($\alpha=\beta$) (dashed lines).
The latter is in agreement with normal band theory calculations \cite{Chamel05} performed with the same mean-field potentials used here.
The superfluid density computed in this way tends to slightly decrease for increasing value of the pairing gap $\Delta$, as also found in \cite{Chamel25}.
As it can be seen, neglecting the contribution of the off-diagonal elements ($\alpha\neq\beta$) to the sum in Eq.~(\ref{eq:supdens}) leads to a strong underestimation of the superfluid density $\rho_S$.
In the $\rho_b=0.03$~fm$^{-3}$ case, adding the contribution of the off-diagonal elements ($\alpha\ne\beta$) results in a two times larger value of $\rho_S$ already at $\Delta\simeq0.01$~MeV, and a more than ten times larger value of $\rho_S$ at $\Delta\simeq 1$~MeV which is probably more realistic. At that value of the gap, the superfluid density is close to the density of dripped neutrons (see appendix).
The behavior we find here is qualitatively in agreement with the one we got in the rod phase \cite{Almirante24A}, where we performed HFB calculations for fixed mean field including the pairing channel self-consistently but varying the strength of the pairing interaction. Here we perform the same analysis for the crystal phase for two values of the coupling constant ($g/g_{\text{phys}}=0.6$ and $0.7$, see \cite{Almirante24A} for details) and show the results in Fig.~\ref{fig:supdens}.

Notice that in Eq.~(\ref{eq:supdens}) we neglected the effect of the perturbation on the pairing gap.
As shown in \cite{Almirante24,Almirante24A}, the perturbation induces a position dependent phase in the gap, very similar to the one obtained within superfluid hydrodynamics \cite{Martin16}.
It is necessary for respecting the continuity equation \cite{Thouless62} and it leads to a second term in the linear response \cite{Migdal59}.
This is probably the reason for the small discrepancy in Fig.~\ref{fig:supdens} between the results of Eq.~(\ref{eq:supdens}) and the HFB results.
As it can be seen, it has only little effect on the superfluid density.
Furthermore, we neglected the density dependent microscopic effective mass $m^*$ since its inclusion requires to add also a current-current interaction to maintain Galilean invariance \cite{Engel75}, adding more terms to the expression of the current \cite{Chamel19} and of the linear response.
These effects are taken into account in our fully self-consistent HFB calculations, building up on our previous works \cite{Almirante24,Almirante24A}, which will be published in a separate paper.

\textit{Conclusions}---To summarize, in this letter we derived the expression for the superfluid density in the BCS approximation. 
The starting point is the linear response of the density matrix \cite{Migdal59} to a perturbation $V=-\vv\cdot\pv$, where $\vv$ is the velocity of the normal fluid.
This is computed in the framework of band theory under the assumption that the pairing gap is constant (BCS approximation) and unaffected by the perturbation.
The perturbation induces a flow from which the superfluid density can be computed. 
We found that the actual expression for the superfluid density contains the contribution already computed in \cite{Carter05A,Chamel25}, plus another contribution resulting from the off-diagonal elements of the density matrix perturbative correction.

With two numerical examples corresponding to average baryon densities of $0.03$ and $0.055$ fm$^{-3}$, we showed in Fig.~\ref{fig:supdens} that the contribution neglected in \cite{Carter05A,Chamel25} can be huge.
Our results support our previous findings obtained through fully self-consistent HFB calculations in the pasta phases \cite{Almirante24,Almirante24A}, which gave superfluid fractions comparable to the ones predicted by superfluid hydrodynamics \cite{Martin16}.

In the literature, the fact that HFB theory may be needed to get the right superfluid density was discussed in \cite{Minami22}, in the framework of a toy model.
Moreover, in the context of ultracold atoms, it was pointed out that the (time-dependent) HFB equations are able to correctly reproduce the hydrodynamical behavior if the coherence length of the Cooper pairs is small compared to the typical length scales of variation of the applied potentials \cite{Grasso05,Tonini06}.
However, in the study of rotating nuclei, it was found that already the current computed in linear response, even if built on top of the BCS approximation, approaches the ideal-fluid behavior in that limit \cite{Migdal59,Thouless62} (cf. also \cite{Urban03} for the case of cold-atoms).
If the pairing gap is much smaller than the depth of the mean-field potential, as is the case in the inner crust of neutron stars, one could expect that the BCS approximation is still valid. Hence, the problem in \cite{Carter05A,Chamel25} is not the BCS approximation for the ground state, but the neglect of the off-diagonal elements in the linear response of the density matrix. 
The need for this additional contribution to get the correct superfluid density, and other related quantities, has also been pointed out in the context of ultra-cold atoms \cite{Peotta15,Liang17} and multiband BCS superconductors \cite{Iskin24,Jiang24,Iskin25}, where it is called ``geometric contribution''. 

As a consequence, we find that the dependence of the superfluid density on the value of the pairing gap is much stronger than previously expected, especially if the gap is small. This means that a reliable determination of the gap is needed to make precise statements about superfluidity in the inner crust of neutron stars.

In conclusion, it is very likely that the entrainment effect in the inner crust is much weaker than commonly thought, and the superfluid density in the inner crust is probably close to $90\%$ of the average neutron density in layers where previous calculations \cite{Chamel12A} predicted a superfluid fraction of less than 10\%.
Even if the gap was reduced by a factor of two or three due to screening effects, it would not drastically change our conclusion.
It might be also worthwhile looking at the effect of disorder, although a previous study \cite{Sauls20} suggests that it would not lead to a substantial reduction of the superfluid fraction.

As it was shown in \cite{Martin16}, with superfluid fractions as we found in this letter, the glitches of the Vela pulsar can be explained with the superfluidity of the crust alone, and it is not necessary to modify the glitch models to include superfluidity in the core as suggested in \cite{Andersson12}.
Our result has also implications for the low-lying phonons \cite{Pethick10,Chamel13B,Durel18} and hence for the thermal evolution of neutron-stars \cite{Page12}, as well as for the frequencies of star oscillation modes \cite{Andersson02,Sotani12,Tews17}.

\begin{acknowledgments}
\textit{Acknowledgments---}We gratefully acknowledge support from the CNRS/IN2P3 Computing Center (Lyon - France) for providing computing resources needed for this work. We thank M. Iskin for drawing our attention to the relationship between our formula and the geometric contribution in condensed matter physics.
\end{acknowledgments}

\bibliography{refs}
\appendix
\section{Appendix}
\textit{Simple derivation of Migdal's formula (\ref{eq:migdal})---}The unperturbed HFB matrix $\calH^{(0)}$ has eigenvalues $E_\alpha$ and $-E_\alpha$. We denote the corresponding unperturbed eigenvectors with positive and negative energies by $|\alpha_+\rangle$ and {$|\alpha_-\rangle$, respectively, and they are given by
\begin{equation}\label{eq:defA}
    |\alpha_+\rangle = 
    \begin{pmatrix}u_\alpha\\v_\alpha\end{pmatrix}
    |\alpha\rangle\,,\quad
    |\alpha_-\rangle = 
    \begin{pmatrix} v_\alpha\\-u_\alpha\end{pmatrix}
    |\alpha\rangle\,.
\end{equation}
The perturbed HFB matrix is written as $\calH^{(0)}+\mathcal{V}$, with
\begin{equation}
    \mathcal{V} = \begin{pmatrix}V& 0\\0& -\bar{V}\end{pmatrix}\,,
\end{equation}
where the matrix elements of $\bar{V}$ are those of $V$ for the time-reversed states (e.g., in coordinate-space representation $\bar{V}=V^*$ \cite{Migdal59}, which amounts to changing $\pv\to-\pv$), such that in our case $\bar{V}=-V$ [cf. Eq.~(\ref{eq:BCSH})]. We write the (negative-energy) perturbed eigenvector in the form $|\alpha_-\rangle^{(V)} = |\alpha_-\rangle+|\alpha_-\rangle'$. Keeping in mind that the complete set of eigenstates contains the eigenvectors to positive and negative energies, the first-order correction $|\alpha_-\rangle'$ can be computed in perturbation theory \cite{Sakurai} as
\begin{align}
    |\alpha_-\rangle' =& \sum_{\beta\ne\alpha} 
    |\beta_-\rangle 
    \frac{\langle \beta_-| \mathcal{V} |\alpha_-\rangle}{-E_\alpha+E_\beta}
    +\sum_{\beta}
    |\beta_+\rangle
    \frac{\langle \beta_+| \mathcal{V} |\alpha_-\rangle}{-E_\alpha-E_\beta}\nonumber\\
    =& \sum_{\beta\ne\alpha} \left[
    \begin{pmatrix}v_\beta \\ -u_\beta \end{pmatrix}
    \frac{v_\beta v_\alpha+u_\beta u_\alpha }{-E_\alpha+E_\beta}\right.\nonumber\\
    &
    +\left.\begin{pmatrix}u_\beta \\ v_\beta \end{pmatrix}
    \frac{u_\beta v_\alpha-v_\beta u_\alpha}{-E_\alpha-E_\beta}\right]V_{\beta\alpha}|\beta\rangle\,,\label{eq:correctionA}
\end{align}
where $V_{\beta\alpha}=\langle\beta|V|\alpha\rangle$. Notice that in the last line, we have written $\beta\ne\alpha$ also in the second sum since the term with $\beta=\alpha$ is zero. The generalized density matrix is defined as \cite{Almirante24}
\begin{equation}
    \mathcal{R} =\sum_\alpha |\alpha_-\rangle\langle\alpha_-| = \begin{pmatrix}\rho&\kappa\\ \kappa^\dagger&1-\bar{\rho}\end{pmatrix},
\end{equation}
where $\kappa=\mathcal{R}_{12}$ is the anomalous density matrix and $\rho=\mathcal{R}_{11}$ is the density matrix.
One should sum over all states with negative perturbed energy, but due to the finite gap, a small perturbation cannot change the signs of the eigenvalues. 
Writing $\mathcal{R}^{(V)} = \mathcal{R}+\mathcal{R}'$, and keeping only the linear terms, the correction to the generalized density matrix is given by
\begin{equation}
    \mathcal{R}' = \sum_\alpha (|\alpha_-\rangle'\langle\alpha_-|
    +|\alpha_-\rangle\langle\alpha_-|')\,.
\end{equation}
Using Eqs.~\eqref{eq:defA} and \eqref{eq:correctionA} and the hermiticity of $V$ (i.e., $V_{\alpha\beta}=V_{\beta\alpha}$), one finds
\begin{equation}
    \rho'_{\alpha\beta} =
    \langle\alpha|\mathcal{R}_{11}|\beta\rangle\\
    = -\frac{(v_\alpha u_\beta-u_\alpha v_\beta)^2}{E_\alpha+E_\beta}V_{\alpha\beta}\,. 
\end{equation}
Finally, using Eq.~\eqref{eq:BCSEv}, one obtains Eq.~\eqref{eq:migdal}.

\textit{Proof of Eq.~(\ref{eq:responseR-final})---}Inserting Eq.~\eqref{eq:momentum} into Eq.~\eqref{eq:responseR-rewritten}, and using $\partial/\partial k^i=(\partial \xi_\alpha/\partial k^i) \partial/\partial \xi_\alpha$, one obtains
\begin{align}
    \frac{R^{ij}}{m} =& -2\int_{\BZ}\frac{d^3k}{(2\pi)^3} \Bigg[
    \sum_\alpha \frac{\partial v_\alpha^2}{\partial k^i} p^j_{\alpha\alpha}\nonumber\\
    &+\sum_{\alpha\ne\beta,\nv}(v_\alpha^2-v_\beta^2)
    \frac{\partial\langle\alpha|\nv\rangle}{\partial k^i}\langle\nv|\beta\rangle
    p^j_{\beta\alpha}\Bigg].
\end{align}
In the second term, the restriction $\alpha\ne\beta$ can be omitted because the diagonal contributions are anyway zero. Then, renaming in the $v_\beta^2$ term $\alpha\leftrightarrow\beta$ and using \begin{equation}\label{eq:derivative-property}
  \sum_{\nv}\frac{\partial\langle\beta|\nv\rangle}{\partial k^i}\langle\nv|\alpha\rangle = -\sum_{\nv}\langle\beta|\nv\rangle\frac{\partial\langle\nv|\alpha\rangle}{\partial k^i}\,,
\end{equation}
one gets
\begin{align}
    \frac{R^{ij}}{m} =& -2\int_{\BZ}\frac{d^3k}{(2\pi)^3} 
    \Bigg[
    \sum_{\alpha} 
    \frac{\partial v_\alpha^2}{\partial k^i} 
    p^j_{\alpha\alpha}\nonumber\\
    &+\sum_{\alpha,\beta,\nv} v_\alpha^2
    \Big(
    \frac{\partial\langle\alpha|\nv\rangle}{\partial k^i}
    \langle\nv|\beta\rangle p^j_{\beta\alpha}
    +\langle\beta|\nv\rangle
    \frac{\partial\langle\nv|\alpha\rangle}{\partial k^i}
    p^j_{\alpha\beta}
    \Big)\Bigg].
\end{align}
Now, using in the second term the completeness $\sum_\beta |\beta\rangle\langle\beta|=1$, inserting in the first term a complete set of momentum eigenstates $\sum_{\nv} |\nv\rangle\langle\nv|=1$, and using the eigenvalue of the momentum operator $p^j|\nv\rangle = (n^j+k^j)|\nv\rangle$:
\begin{align}
    \frac{R^{ij}}{m} =& -2\int_{\BZ}\frac{d^3k}{(2\pi)^3} 
    \sum_{\alpha,\nv}(n^j+k^j)\Bigg[
    \frac{\partial v_\alpha^2}{\partial k^i} 
    |\langle\alpha|\nv\rangle|^2\nonumber\\
    &+v_\alpha^2\Big(
    \frac{\partial\langle\alpha|\nv\rangle}{\partial k^i}
    \langle\nv|\alpha\rangle
    +\langle\alpha|\nv\rangle
    \frac{\partial\langle\nv|\alpha\rangle}{\partial k^i}
    \Big)\Bigg]\nonumber\\
    =&-2\int_{\BZ}\frac{d^3k}{(2\pi)^3} 
    \sum_{\nv\alpha}(n^j+k^j)
    \frac{\partial (v_\alpha^2|\langle\alpha|\nv\rangle|^2)}{\partial k^i}\,.
   \label{eq:responseRd3k}
\end{align}
For clarity, we will now write the labels $\kv$ that we usually omit. Integrating Eq.~\eqref{eq:responseRd3k} by parts, we get
\begin{align}
  \frac{R^{ij}}{m} =& 2\int_{\BZ}\frac{d^3k}{(2\pi)^3} \sum_\alpha v_{\alpha\kv}^2\,\delta^{ij}\nonumber\\
  &-\frac{1}{4\pi^3}\int_{\partial\BZ}dS^i
  \sum_{\alpha\nv}
  (n^j+k^j) v_{\alpha\kv}^2
  |\langle \nv\kv|\alpha \kv\rangle|^2\,,
   \label{eq:responseRdS}
\end{align}
where the second term is an integral over the surface $\partial\BZ$ of the first BZ, the vector $d\vect{S}$ pointing outwards. 
For each point $\kv \in \partial\BZ$ on one face of the BZ, there is a point on the opposite face, i.e., $\kv'\in \partial\BZ$ with $d\vect{S}'=-d\vect{S}$, which can be reached by making a translation of the integer $\nv$ labels such that $\nv'+\kv' = \nv+\kv$ (in the case of a simple cubic lattice, it is enough to change one of the three $n$ labels by $\pm 1$).
In other words, $|\nv'\kv'\rangle = |\nv\kv\rangle$.
As a consequence, for Bloch momentum $\kv'$ there exists an eigenstate $|\alpha'\kv'\rangle$ with energy $\xi_{\alpha'} = \xi_\alpha$ and hence $v^2_{\alpha'} = v^2_\alpha$, whose matrix elements satisfy $\langle\nv'\kv'|\alpha'\kv'\rangle = \langle\nv\kv|\alpha\kv\rangle$.
Therefore, when summed over all states, the surface integral in Eq.~\eqref{eq:responseRdS} cancels, which proves Eq.~\eqref{eq:responseR-final}.

\textit{Dripped neutrons and macroscopic effective mass---}
In the literature, entrainment has often been described in terms of the macroscopic effective mass $m_\star$ \cite{Chamel05}.
To compute this quantity, one needs to define the density of dripped (or ``free'') neutrons $\rho_f$.
Then, one can get the macroscopic effective mass $m_\star$ as \cite{Chamel05}
\begin{equation} \label{eq:effective_mass}
    \frac{m_\star}{m}=\frac{\rho_f}{\rho_S} \,.
\end{equation}
%
\begin{table} [h]
  \caption{\label{tab:effective_mass} Results for the macroscopic effective mass $m_\star$ from Eq.~(\ref{eq:effective_mass}), using the density of dripped neutron $\rho_f$ from \cite{Chamel05} and the superfluid density $\rho_S$ from Eq.~(\ref{eq:supdens}) evaluated at $\Delta=1$ MeV.}
  \begin{ruledtabular}
    \begin{tabular}{cccc}
        $\rho_b$~(fm$^{-3})$&$\rho_f^{\text{\cite{Chamel05}}}/\langle\rho\rangle$&$m_\star^{\text{\cite{Chamel05}}}/m$&$m_\star/m$\\\hline
        0.030&0.94&15.4&1.06\\
        0.055&0.95&5.5&1.02
    \end{tabular}
  \end{ruledtabular}
\end{table}
However $\rho_f$ is not uniquely defined, while $\rho_S$ is \cite{Carter06,Pethick10}.
Still, since we use the same mean-field potentials as in \cite{Chamel05} and for the sake of comparison, here we compute the macroscopic effective mass $m_\star$ using the values for the density of dripped neutrons $\rho_f$ given in \cite{Chamel05}. 
In Table~\ref{tab:effective_mass} we show results for $m_\star$ computed with $\rho_S$ in Eq.~(\ref{eq:supdens}) and compared with results from \cite{Chamel05}. 
As it can be seen, at $\Delta=1$ MeV, our result for the macroscopic effective mass $m_\star$ is one order of magnitude smaller than predicted in $\cite{Chamel05}$ for the case were entrainment was the strongest, and for both the explored cases it is very close to the bare mass $m$, indicating that entrainment is practically negligible.

\end{document}